\documentclass[journal]{IEEEtran}

\usepackage{cite}

\usepackage{mwe}
\usepackage{epsfig}
\usepackage{epsf}
\usepackage{epstopdf}
\usepackage{times}
\usepackage{float}
\usepackage{amssymb}
\usepackage{amsmath}

\DeclareMathOperator*{\argmin}{arg\,min}
\usepackage{cite}
\usepackage{color}
\usepackage{tabularx}
\usepackage{acronym}
\usepackage{mathtools}
\usepackage{booktabs}
\usepackage{makecell}
\usepackage{hyperref}

\usepackage{algorithm}

\usepackage{graphicx,mwe}
\usepackage{caption}
\usepackage{subcaption}
\usepackage{xcolor, soul}
\sethlcolor{red}

\usepackage{soul}
\usepackage{bm}
\usepackage{amsfonts}

\usepackage{dsfont}
\usepackage{lipsum}
\usepackage{booktabs}
\usepackage{nicematrix}

\hypersetup{colorlinks=true,linkcolor=black,anchorcolor=black,citecolor=black,filecolor=black,menucolor=black,runcolor=black,urlcolor=black}

\makeatletter
\newcommand\remembertext[2]{% #1 is a key, #2 is the text
  \immediate\write\@auxout{\unexpanded{\global\long\@namedef{mytext@#1}{{\color{blue} #2}}}}%
  {#2}
}
\newcommand\remembertextnocolor[2]{% #1 is a key, #2 is the text
  \immediate\write\@auxout{\unexpanded{\global\long\@namedef{mytext@#1}{#2}}}%
  #2%
}
\newcommand\recalltext[1]{%
  \ifcsname mytext@#1\endcsname
    \@nameuse{mytext@#1}%
  \else
    ``??''
  \fi
}
\makeatother

\ifCLASSINFOpdf
\else

\fi

\begin{document}

\title{Detecting the Presence of Sperm Whales’ Echolocation Clicks in Noisy Environments}
\author{\IEEEauthorblockN{\large
		Guy~Gubnitky$^{\S}$,
		Roee~Diamant$^{\S,*}$\\[1.5mm]}
\IEEEauthorblockA{\small $^\S$Hatter Department of Marine Technologies, University of Haifa, Israel\\}
\IEEEauthorblockA{\small $^*$Faculty of Electrical Engineering and Computing, University of Zagreb, Croatia\\}%
\IEEEauthorblockA{\small $^\star$Corresponding author, email: {\tt roee.d@univ.haifa.ac.il}%
 \vspace{-6mm}}
 \thanks{Parts of this work have been presented~\cite{gubnitsky2023inter}. This journal version
extends this work by adding detection and verification methods and  results from two state-of-the-art benchmarks.}
\thanks{This research was supported by the TED Audacious project CETI, and by the Horizon Europe programme of the European Union under the UWIN-LABUST project (project number 101086340).}
}

%\markboth{Journal of \LaTeX\ Class Files,~Vol.~13, No.~9, September~2014}%
%{Shell \MakeLowercase{\textit{et al.}}: Bare Demo of IEEEtran.cls for Journals}

%\begin{document}

\maketitle

\begin{abstract}

\remembertext{R1C15c}{Sperm whales (\textit{Physeter macrocephalus}) navigate underwater with a series of impulsive, click-like sounds known as echolocation clicks. These clicks are characterized by a multipulse structure (MPS) that serves as a distinctive pattern. In this work, we use the stability of the MPS as a detection metric for recognizing and classifying the presence of clicks in noisy environments. To distinguish between noise transients and to handle simultaneous emissions from multiple sperm whales, our approach clusters a time series of MPS measures while removing potential clicks that do not fulfil the limits of inter-click interval, duration and spectrum.} As a result, our approach can handle high noise transients and low signal-to-noise ratio. The performance of our detection approach is examined using three datasets: seven months of recordings from the Mediterranean Sea containing manually verified ambient noise; several days of manually labelled data collected from the Dominica Island containing approximately 40,000 clicks from multiple sperm whales; and a dataset from the Bahamas containing 1,203 labelled clicks from a single sperm whale. Comparing with the results of two benchmark detectors, a better trade-off between precision and recall is observed as well as a significant reduction in false detection rates, especially in noisy environments. \remembertext{R1C1a}{To ensure reproducibility, we provide our database of labelled clicks along with our implementation code.}

\end{abstract}


\IEEEpeerreviewmaketitle

\section{Introduction}

\IEEEPARstart{S}{perm Whales} (SWs) have been identified as a species with an advanced non-human communication system due to their highly developed brain structure~\cite{marino2011brain}, their unique form of communication\cite{watkins1977sperm} and their complex multi-level societies~\cite{gero2016individual}. Their vocalizations consist of either a series of stereotyped transient patterns (called codas")~\cite{watkins1977sperm} or impulse-like signals used for echolocation, known as clicks~\cite{tonnesen2020long}. Automatic detection of SWs using passive underwater acoustic monitoring (PAM) is an important tool for studying the presence, distribution, migration and behaviour of SWs. This work is part of the CETI project~\cite{andreas2021cetacean}, in which an interdisciplinary research group has started to collect huge amounts of SW vocalizations to fulfil the ambitious task of deciphering the language of SWs.
This task requires the identification and characterization of codas and clicks. While the former provide information on the whales' social interactions (e.g. identification of clans and social units, communication content and abundance), the latter allow researchers to understand their foraging behaviour, biosonar capabilities, ecology and population sizes. In this paper, we focus on the detection of SW clicks.

\iffalse
\begin{figure}[]
	\centerline{
		\includegraphics[width=90mm,height=5cm]{Figures/head_add2.png}}
	\caption{ Illustration of SW vocalization for codas and echolocation clicks. The time interval between pulses ejected from the phonic lips and distal air sac, $p'_1-p_0$ or $p'_n-p'_{n-1},~n=2,3,...$, is called IPI. The time interval between consecutive echolocation clicks is called ICI. 
}
	\label{fig:12}
\end{figure}
\fi

\remembertext{R1C3a}{One of the biggest challenges in SW detection is the directional nature of the clicks, which leads to angle-dependent distortions of both the waveform and the spectrum in off-axis recordings~\cite{goold1995time}. That is, when the head of the SW is not directed towards the receiver~\cite{zimmer2005off}. Examples of this challenge for groups of 6 SW clicks can be found in Fig.~\ref{fig:High SNR clicks} and in Fig.~\ref{fig:Low SNR clicks} for clicks received with a high SNR of 44 dB and a low SNR of 24 dB, respectively. In both cases, we observe a considerable difference between the temporal and spectral content of the different clicks in the series. This challenge in terms of variability of clicks is compounded by the similarity of click sounds to the sound of noise transients of e.g. snapping shrimps or the cavitation noise of ships. For this reason, human-in-the-loop approaches~\cite{webber2022streamlining,gordon2020first,lewis2018abundance} are still used.

\begin{figure}
\centering  
  \begin{subfigure}[t]{.95\columnwidth}    \centering\includegraphics[width=1\linewidth]{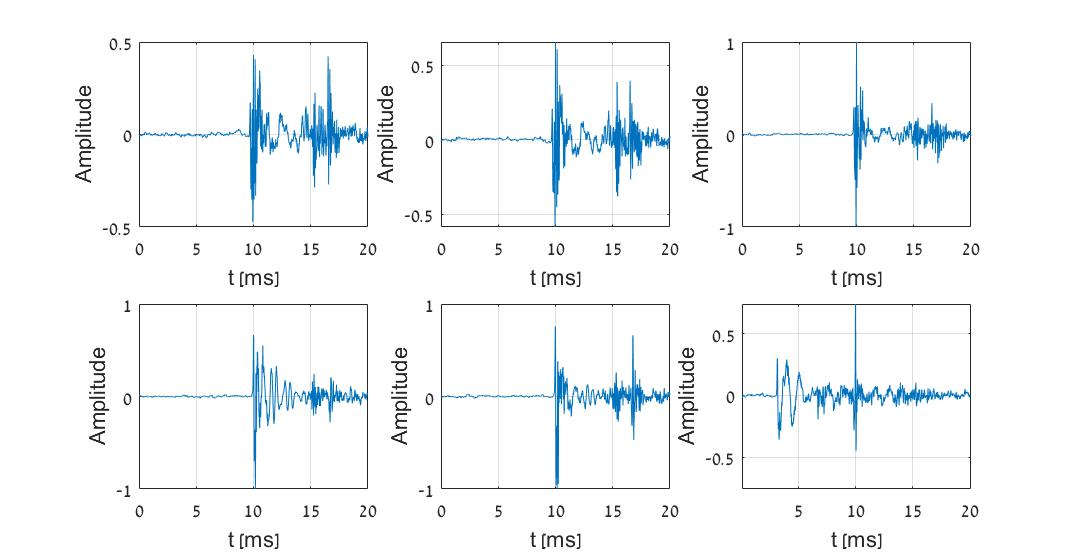}
  \caption{Apparent waveform}
\end{subfigure}
  \begin{subfigure}[t]{.95\columnwidth}    \centering\includegraphics[width=1\linewidth]{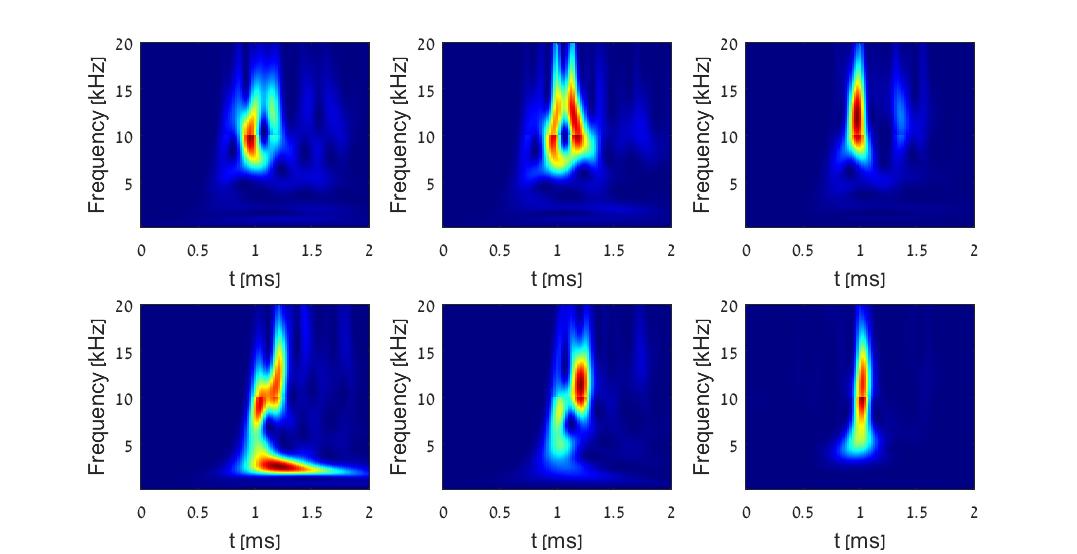}
    \caption{Apparent spectrogram}
\end{subfigure}
\caption{Example of the variability in SWs' clicks. The example demonstrates 6 high SNR click measurements (44dB in average) taken from the AUTEC dataset. }
\label{fig:High SNR clicks}
  \begin{subfigure}[t]{.95\columnwidth}    \centering\includegraphics[width=1\linewidth]{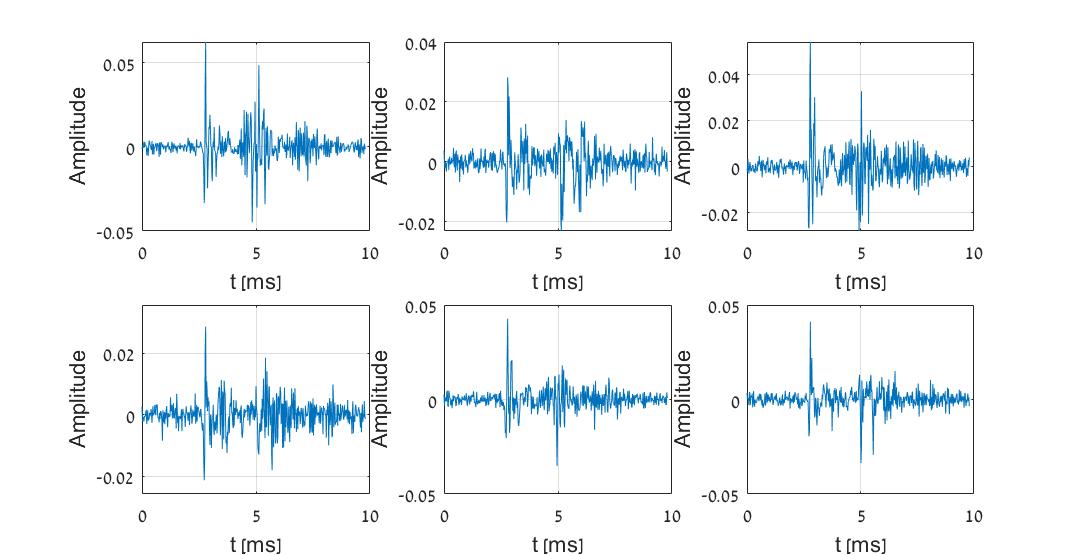}
  \caption{Apparent waveform}
\end{subfigure}
  \begin{subfigure}[t]{.95\columnwidth}    \centering\includegraphics[width=1\linewidth]{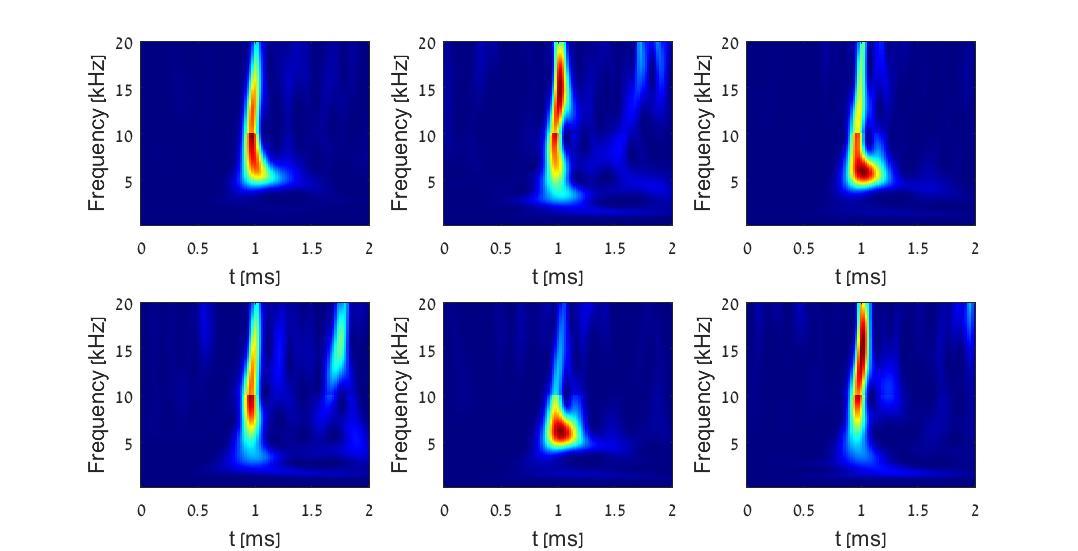}
    \caption{Apparent spectrogram}
\end{subfigure}
\caption{Example of the variability in SWs' clicks. The example demonstrates 6 low SNR click measurements (24dB in average) taken from the Dominica dataset.}
\label{fig:Low SNR clicks}
\end{figure}

Approaches to SW detection are divided into the detection of individual clicks or presence detection within a certain time buffer. Recent schemes proposed supervised machine learning to classify features extracted from the waveform of the signal~\cite{sanchez2010efficient,roch2011classification,van2009classification}, while training is performed over a large, locaiton-specific database~\cite{gibb2019emerging}. Another approach offered is template matching, in which PAM measurements are correlated with a model of a click. This is the method in \cite{caudal2008stochastic}, which performs stochastic matched filtering, and the method in \cite{kandia2006detection}, where the signal is matched filtered with a waveform model template. In \cite{lohrasbipeydeh2014adaptive}, the signal is filtered by an average of the previously detected clicks. Similarly, the normalised average spectrum of clicks within a click sequence is correlated with a species spectrum template.
%~\cite{kandia2006detection}, with the averaged waveform~\cite{lohrasbipeydeh2014adaptive,caudal2008stochastic} or spectrum~\cite{macaulay2020passive,webber2022streamlining} of known click templates acquired from previous recordings.
However, template matching offers only a low generalization capacity, especially in the presence of noise transients (e.g. from ship propellers or snapping shrimps), which are similar in structure to SW clicks.

To reduce false detections, recent works proposed to combine classification with contextual information such as the click rate~\cite{macaulay2020passive,skarsoulis2022real,baggenstoss2011algorithm} and similarity of signal features~\cite{frasier2021machine} for presence detection within a certain time buffer. A strong contextual indicator is the inter-click interval (ICI), which allows the separation of SW clicks from the clicks of other toothed whales~\cite{deangelis2018description,baumann2013species,roberts2015field} and from noise transients, especially when combined with directional analysis using a set of receivers~\cite{skarsoulis2022real}. However, identifying the ICI in a time buffer that contains clicks from multiple SWs in addition to a large number of noise transients is challenging. Some solutions are offered to separate the sources of the clicks~\cite{macaulay2020passive,le2015rhythmic,bahl2003techniques}, but the results are limited to relatively simple cases with high SNR values.}

\begin{figure}    \centering\includegraphics[width=1\columnwidth]{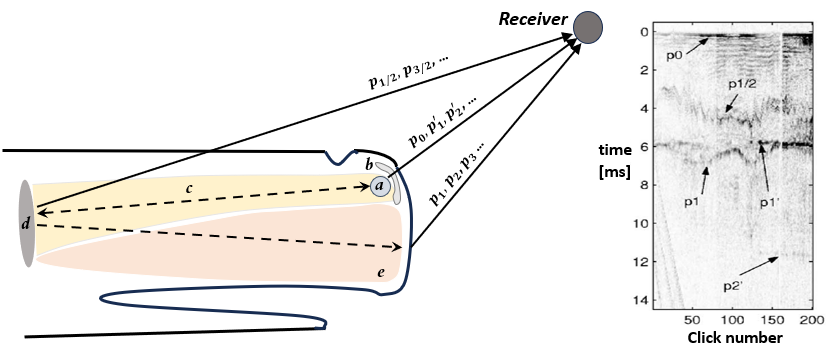}
    \caption{Left: the generation of SW clicks according to the leaky bent-horn model~\cite{laplanche2006measuring}. Right: an example of a multi-pulse layout from measurements of 200 consecutive clicks (taken from~\cite{laplanche2006measuring}).}
\label{MPS demo}
  \end{figure}

Another unique feature of SW clicks is their multi-pulse structure (MPS)~\cite{laplanche2006measuring}, which is a product of internal reflections within the spermaceti organ of the whale - back and forth between the frontal and distal air sacs. As shown in Fig~\ref{MPS demo}, an initial pulse, called $p_0$ is generated at the whale's phonic lips, leading to the emission in each reflection round of three pulses, $p_{1/2}$, $p_1$ and $p_1'$. While the exit point of the pulse $p_1'$ is similar to that of the initial pulse $p_0$ (i.e. the distal air sac), the pulses $p_{1/2}$ and $p_1$ emerge from the frontal sac and the front end of the junk, respectively. This results in a consistent time interval between $p_0$ and $p_1'$, called the inter-pulse interval (IPI), which correlates with the size of the spermaceti of the emitting whale~\cite{gordon1991evaluation,antunes2010measuring}, as well as variable intervals between the other pulses, which are a function of the orientation of the whale in relation to the receiver. \remembertext{R1C15e}{The MPS feature is used in the context of IPI and SW size estimation~\cite{caruso2015size,laplanche2006measuring,gordon1991evaluation,pavan1997software,antunes2010measuring}. However, to our knowledge, this feature has not been used directly for the detection and classification of SW clicks.}

In this work, we present a method for detecting SW presence, which we call MPS-based clicks detection (\textit{MPS-CD}). Our solution aims to detect the presence of a SW within a time buffer (10~sec in our results) as a first, low-complexity step towards more accurate detection using arrays of receivers. \remembertext{R1C15d}{Our scheme uses the stability of the MPS as the detection metric. Specifically, for a given time buffer with possible clicks and corresponding MPS values, we search whether there is a set of stable MPSs that fulfil constraints on click duration, ICI and spectra. Using clustering, we overcome the challenge of noise transients that resemble SW clicks and different MPS values emanating from multiple whales within the time buffer. Our contribution is twofold: \begin{itemize}
\item A detection scheme to recognize the presence of SWs based on the stability of the MPS of the clicks.
\item An algorithm for clustering SW clicks that is robust to the number of sources and to noise transients.
\end{itemize}
}
\remembertext{R1C4a}{We analyze the results of three datasets: 1) seven months of recordings from the THEMO observatory~\cite{diamant2018themo}, containing both ambient and anthropogenic noise but no whale clicks, 2) recordings of 1203 tagged clicks of a single whale from the Bahamas with 120 sec of noise-only recording and 3) recordings of over 15,000 manually tagged clicks from multiple whales we recorded on the Dominica Island together with 3.6~hours of noise-only data. The performance is analyzed in terms of the precision-recall tradeoff and the false alarm rate to test the robustness for the marine environment. The results show a significant improvement compared to two benchmark methods in the case of noisy environment.} Our proposed method has been implemented in real-time and is currently being used for in-situ detection of SW vocalizations on the Dominica Island as part of the CETI project. \remembertext{R1C2}{The implementation code of our detector and our dataset for both SW clicks and noise transients, as well as our manual annotations are available in~\cite{Dominica}.}

The remainder of this paper is organized as follows. In Section~\ref{survey}, we give an overview of the state of the art in click detection. Section~\ref{system model} contains our system model and the basic assumptions. A description of our method can be found in Section~\ref{method description}. Performance and discussion are presented in Section~\ref{results}. Conclusions are drawn in Section~\ref{sec:conclude}. Table~\ref{table: list of variables} outlines the main notations of the paper.

% \begin{table*}
% \caption{List of parameters} % title of Table 
% \centering      % used for centering table
% \begin{tabular}{c c }  % centered columns(4 columns) 
% \hline\hline  %inserts double horizontal lines 
%  Parameter  & Description  \\ [0.5ex] % inserts table 
% %heading 
% \hline  
%            $\zeta$   & Tolerance parameter to off-axis excitations     \\ 
%            $t_{\mathrm{buf}}$  & Duration of the given time buffer  \\ 
%            $N$       &  Number of signal samples within the given time buffer    \\  
%            $\mathcal{N}$   &  Number of simulated noise IPI-sets     \\
%            $\mathcal{M}$   &  Number of simulated noise TOA-sets     \\
%            $L_{\mathrm{max}}$  & maximum number of transients considered for a time buffer     \\
%            $\eta$   &  The entire noise area in the feature space    \\
%            $\eta^u$   &  The unsegmented noise region in the feature space   \\
%            $\eta^s$   &  The segmented noise area in the feature space     \\
%            $T_{\mathcal{C}}$   &  A constraint for clustering - a bound over the maximal consistency measure  \\
%            $T_{\mathcal{J}}$   &  A constraint for clustering - a bound over the maximal variability in the signals' intensity   \\
%             $T_{\mathrm{ICI}}$   &  Verification threshold - a bound over the maximum CTR    \\
%             $\Upsilon$   &  Detection threshold- a bound over the maximum number of noise transients remaining after segmentation     \\           
% \hline     
% \end{tabular} 
% \label{table: list of parameters}  
% \end{table*}  

\begin{table}
    
\caption{List of main variables} % title of Table 
\centering      % used for centering table
\begin{tabular}{c c }  % centered columns(4 columns) 
\hline\hline  %inserts double horizontal lines 
 Variable  &  Description   \\ [0.5ex] % inserts table 
%heading 
\hline  

           y          & measured signal       \\
           s          & enhanced signal          \\
            $N$       &  number of signal samples within the given time buffer    \\
            ICI     & inter-click interval \\
            IPI     & inter-pulse interval \\
            MPS     & multi-pulse structure \\
           $\rho_{\mathrm{MPS}}$     & click's MPS measurement    \\    
           $\mathcal{C}$  & Consistency of clicks' ICI  \\
           $\rho_d$  &  pulse duration   \\
           $\rho_f$  &  pulse resonant frequency   \\
        
\hline     
\end{tabular} 
\label{table: list of variables}  
\end{table}  

\section{Related work}
\label{survey}

Detection of SW vocalization by template matching requires a model of the click waveform~\cite{kandia2006detection} or comparison with previously measured clicks~\cite{lohrasbipeydeh2014adaptive}. Supervised machine learning is an alternative if the model does not mismatched or if the noise transients are not too similar to the template used. For example, in \cite{bermant2019deep}, a convolutional neural network (CNN) is used to classify spectrograms for SW presence detection, while \cite{sanchez2010efficient} involves a preliminary feature extraction step followed by a multilayer perceptron (MLP) network to assign the time windows to SW or noise classes. A feature-based classifier based on \cite{van2009classification} is proposed, where a support vector machine (SVM) with a radial basis function is used to classify features represented by wavelet packets of clicks. This wavelet transform is based on the choice of vectors for the signal decomposition (i.e. the wavelet family), which is signal-dependent. A possible alternative is the Hilbert-Huang transform (HHT)~\cite{adam2006use}. Unlike the Fourier and wavelet transforms, the HHT does not use specific vectors for the decomposition, but decomposes the signal into itself.

One class of supervised approaches to SW detection is centred on classification, where a soft decision is first applied to detect transients and then SW clicks are identified from a set of noise categories. For example, in~\cite{zaugg2010real} impulsive sounds with a desired frequency content are recognized and classified as SW clicks or ship impulses. Similarly, in~\cite{roch2011classification} the echolocation clicks of 6 odontocetes species are first detected using energy and spectral criteria. Then a set of cepstral features is extracted and classified using Gaussian mixture models (GMM). However, the classification approach assumes a certain number of noise sources, the number of which is unknown in most PAM applications. In addition, the above two supervised detection and classification methods rely on the waveform features, whose structure may change due to temporal effects in the water medium, so unsupervised approaches can sometimes provide better results~\cite{lohrasbipeydeh2014adaptive}.

In cases where no template or trained model is available, recognition with unsupervised approaches is possible. In~\cite{walker2003automated}, a change detector is proposed for finding a transient click, where changes in signal intensity are identified by counting the number of occurrences where the change in intensity level exceeds a threshold. A recursive detection technique~\cite{lopatka2006sperm} builds a lattice-based filter called a Schur filter, to detect transients. At each time step, the adaptive filter calculates an optimal orthogonal signal representation using the second-order statistics of the signal. The result is a series of time-varying filter coefficients that highlights non-stationary transient events. To account for the high acoustic intensity of clicks over a short period of time~\cite{levenson1974source}, some unsupervised methods for click detection are based on the concept of energy detection. One of these methods is the Teager-Kaiser energy operator (TKEO), which utilizes both amplitude and frequency information~\cite{kaiser1990simple}. Based on the TKEO, an energy ratio mapping algorithm is presented in~\cite{klinck2011energy} to cluster the desired and undesired transients and calculate the energy ratios between the two. However, this method is not robust for the case where the energy fluctuates due to the frequent orientation changes of the SW. To overcome the difficulties in setting an appropriate detection threshold, an adaptive TKEO (ATEO) is proposed~\cite{lohrasbipeydeh2014adaptive} to update the threshold based on the mean and variance of the TKEO output. Another approach is to approximate the TKEO output by a Gaussian function~\cite{madhusudhana2015automatic} followed by a spike detection scheme that incorporates a scaled Gausian moving- average filter.

\remembertext{R1C3c}{Another type of detector is context-based, which is based on the periodicity of the ICI of the SW clicks. In~\cite{frasier2021machine}, a two-stage unsupervized clustering algorithm is used to discard non-recurring signal types based on their similarity. Other approaches focus on the unique, slowly varying ICI characteristics of the click sequences of SWs. In \cite{skarsoulis2022real}, a detection algorithm is provided that measures the distribution of arrival time differences between three distant hydrophones to search for peaks within a known range. In~\cite{macaulay2020passive}, a multi-hypothesis tracking (MHT) is developed to detect click sequences of SWs. First, a matrix containing sets of possible click associations is formed. Then, a likelihood measure is calculated for each of the click combinations assuming the statistical metric $\chi^2$ for the slowly varying features of the click sequences. However, the performance strongly depends on the availability of bearing information. Few works have proposed to separate click sequences from simultaneously clicking whales based on rhythmic analysis~\cite{le2015rhythmic}, propagation path~\cite{baggenstoss2011separation} or spectral dissimilarity between closely spaced clicks~\cite{zaugg2013extraction}. However, many non-SW clicks fulfil the similarity criteria and sensitivity to noise transients is observed.}

\remembertext{R1C3b}{In summary, existing approaches to SW click detection rely on classification based on signal statistics or contextual information from click series and require calibration of parameters or training with manually labelled clicks from the measurement site. While the problem of SW detection and in particular presence detection is well studied, we note a knowledge gap in terms of robustness of the methods in different marine environments and in terms of sensitivity to noise transients, which are similar in structure to SW clicks. Considering this, we propose an unsupervised solution that is robust to the environment and measurement equipment and can manage a high rate of noise transients with only a single hydrophone.}

\section{System Model}
\label{system model}

\subsection{Setup of Data Collection}

% \begin{figure}[t]
% 	\centerline{
% 		\includegraphics[width=90mm,angle=0]{Figures/Context.png}}
% 	\caption{The PAM system of CETI. Three phases are considered: detection, localization and tracking of echolocation and coda type signals. This work is focused on the detection of SW echolocation clicks (marked in orange).}
% 	\label{fig:2}
% \end{figure}

Our SW click detection algorithm is specifically designed for  project CETI, which requires on-site analysis of a large data collection of clicks for SW behaviour analysis. At a lower level, this task requires the detection of SW vocalizations, and the estimation of their arrival time to localize the source. Project CETI system comprizes three tethered moorings, each containing a surface buoy connected via an optical fiber to 5 Whale Recording Units (WRU) stationed along the water column from 40~m depth to 800~m depth. Each WRU contains 4 or 12 Colmar-manufactured hydrophones along with temperature and depth sensors. The acoustic data from each hydrophone is sampled synchronously by an RTSys sampling card (sampling at 156~kHz, 3~bytes per sample), which also provides the orientation of the WRU with respect to magnetic north. The data from all WRUs is then streamed to an NVEDIA Jetson board located in the surface buoy. The data is stored for offline analysis (once a month) as well as processed online to provide information on the position of the whales. %An illustration of the system can be found in Fig.~\ref{Ceti mooring}.
So far, a single mooring has been installed, with two more planned for the coming months.

Our detector runs over a time buffer of acoustic data from a single channel. Our MPS-CD approach is based on the identification of the MPS of the clicks. Therefore, we limit our work to the detection of signals from the far field (i.e., from a distance of at least 5 wavelengths) where the inter-pulses can be easily separated. We assume that the SW clicks are mainly in the frequency band $2-24$kHz \cite{lohrasbipeydeh2014adaptive}, and we consider the case of simultaneously emitting whales.

\subsection{Preliminaries}
\label{sec: preliminaries}

%The generation of SW clicks is explained by the leaky bent-horn model~\cite{laplanche2006measuring}. According to this model, clicks are generated in the SW's snout by its phonic lips. A part of the initial pulse, $p_0$, is directed backwards by a distal air sac, passes through the spermaceti case, and is reflected again by a frontal air sac. A part of the frontal reflected pulse is then ejected out of the whale's body as an omnidirectional pulse termed $p_{1/2}$, and part is projected through the spermaceti and focused by its lens-like structure. The signal is then ejected as a powerful and highly directed pulse called $p_1$. While the time lag $p_{1/2}-p_0$  is dictated by the whale’s orientation with respect to the receiver, the time lag $p_1-p_0$ is aspect-independent since both $p_0$ and $p_1$ pulses are ejected from the same air sac. Therefore, $p_1-p_0$ defines the IPI and is often used to estimate the whale’s size~\cite{laplanche2006measuring}.  The click generation process is illustrated in Fig.\ref{fig:1}  %(adapted from \cite{andreas2021cetacean}\cite{laplanche2006measuring}).

The following refers to the illustration in Fig.~\ref{MPS demo}, which describes the generation of SW clicks and shows a multi-pulsed layout from measurements of 200 consecutive clicks. The generation of SW clicks is explained by the leaky bent-horn model~\cite{laplanche2006measuring}. The model distinguishes between 3 types of pulse series.
Type 1 considers pulses that are generated by the phonic lips ($p_0$) that bounce back and forth through the spermaceti case between the frontal and distal air sacs. At each reflection round, part of the energy of the pulse leaves the whale from the distal sac ($p_1', p_2',p_3',...$). Type 2 takes into account the 'half pulses' labelled $p_{1/2}, p_{3/2},...$ that exit the frontal sac. Type 3 considers pulses reflected from the frontal sac and channelled and projected forward by the spermaceti junk, resulting in the outgoing pulse series $p_1, p_2, p_3,\ldots$. Since the generation and exit points of the Type~1 pulses are adjacent, the time lag between these pulses, namely $p_1'-p_0, p_2'-p_1',\dots$, is maintained regardless of the angle of the receiver with respect to the axis of the whale. These time delays are called inter-pulse intervals (IPIs) and are correlated with the length of the whale~\cite{goold1996signal},\cite{rhinelander2004measuring}. In contrast, for the other two types, the origin and exit points of the pulse are far apart, so the measured time difference between them varies with the orientation of the whale relative to the receiver. Note that these variations are more profound with Type~2 pulses, as the distance between the exit point and the phonic lips is greater. These pulses are demonstrated in Fig.~\ref{MPS demo}.

The SW is considered the most acoustically active toothed whale. It navigates and forages by vocalizing echolocation clicks, a series of short and highly directive pulses emitted at a constant time interval (ICI). The ICI characterizes three main phases of diving~\cite{miller2004sperm}: (1) a vertical descent accompanied by a series of clicks at a slow rate: $0.45<\mathrm{ICI} < 2$~sec (\textit{usual clicks}); (2) a hunting phase which takes place in deep water and is characterized by a regular ICI sequence followed by clicks with increasing repetition rate $0.01<\mathrm{ICI} < 0.45$~sec (\textit{creaks}); and (3) a reascent phase: at the end of the hunting phase, the whale returns to the surface to breathe, sleep and communicate with other whales. Communication involves a different type of click called \textit{coda}. In contrast to echolocation clicks, codas are much less directional and consist of lower spectral components and a longer click duration. The ICI model of codas is not consistent, but is composed of various unique stereotyped sequences of 3-40 clicks, usually in the range of $0.01<\mathrm{ICI} < 0.45$~sec. %Each sequence reflects a unique social functionality~\cite{gero2016individual}.

\begin{figure*}[]
 	\centerline{
 		\includegraphics[width=160mm,angle=0]{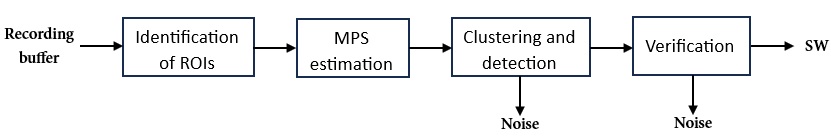}}
 	\caption{ Block diagram of the proposed MPS-CD approach. }
 	\label{Block diagram}
 \end{figure*}

\subsection{Main Assumptions}

\remembertext{R1C8a}{We assume that the SW generates a consistent, multipulsed click pattern within a certain short time buffer (say 10~sec). These pulses are a product of internal reflections in the whale's organs on a scale of 2-8ms~\cite{rhinelander2004measuring}, and are detected at the receiver either by direct passage of the channel or by reflections from the sea surface. In both cases, it is assumed that the arrival times of the dominant pulses are stable across the time buffer. We base this assumption on previous observations that have shown that the orientation (heading, pitch and yaw) of SWs changes only slowly, especially during diving and approach~\cite{laplanche2006measuring}, and because the MPS is strongly correlated with the off-axis angle of the whale~\cite{zimmer2005off,laplanche2006measuring}. In contrast, the pulses related to noise transients are assumed to be unstable due to their randomness. Therefore, the stability of the MPS can be used to distinguish between clicks and noise transients}

We make a further assumption regarding the consistency of the ICI of the clicks. The work in~\cite{baggenstoss2011separation} has shown that SWs produce echolocation clicks with consistent ICI patterns, and we define their consistency by 
\begin{equation}
	\begin{aligned}
		\mathcal{C}_{\mathrm{ICI}}^{i,j,k}= log\left( \frac{t_k-t_j}{t_j-t_i}\right) \;,
		\label{eq 1}
	\end{aligned}
\end{equation} 
where $t_i$, $t_j$ and $t_k$ are the indices of the time points of three consecutive clicks ($k=j+1, j=i+1$). We model the distribution of $\mathcal{C}_{\mathrm{ICI}}^{i,j,k}$ using a Gaussian mixture model (GMM) with two classes: one for the consistency of the central lobe and one for the outliers. Experimental results in \cite{baggenstoss2011separation} showed that $C_{ICI}^m\sim N(0,0.0526^2)$ and $C_{ICI}^o\sim N(0,0.1534^2)$, where $C_{ICI}^m$ denotes the main lobe of the ICI distribution (where $m$ stands for \textit{main}) and $C_{ICI}^o$ denotes the outliers of the ICI distribution (where $o$ stands for \textit{outliers}). It is assumed that the distribution of outliers is the product of the resolution adjustment in which the SW changes the ICI of its click sequence.

\section{Method Description}
\label{method description}

\subsection{Key Idea}

\remembertext{R1C15f}{Our MPS-CD presence detection is specifically designed for cases where a single time buffer contains clicks from multiple SWs and multiple noise transients. We use MPS stability as a context-based information source.
%In this work, we focus on the stability of clicks' MPS and intensity. We argue that the stability of the time intervals between pulses that make up SW clicks, do not depend on the whale's distance to the receiver, but only on the whale's mobility and size.  Hence, we propose using the above stability metric to provide context-based information gain with the aim of overcoming classification challenges associated with waveform distortions caused by click propagation and directional effects. The \textit{MPS-CD} approach is therefore designed for resilience to conditions of low SNR and excess of noise transients.
The steps of MPS-CD are described in the block diagram in Fig. \ref{Block diagram}. We start with a sampled time buffer $y_n, \quad n=1,\ldots,N$, which is processed to identify regions of interest (ROIs) around dominant transients. We then compute the MPS of each detected transient within the ROI. The series of MPS is clustered to find groups of consistent MPS. This is followed by a verification step to eliminate clicks that do not match the ICI, duration and spectrum constraints. Detection is declared when a valid cluster of MPS values is considered stable enough.}

\remembertext{R1C15a}{The potential in exploring the stability of MPS is demonstrated by the histograms in Fig.~\ref{validation of MPS stability}, where we measure the standard deviation of MPS for groups of five clicks for a database of 1203 manually verified SW clicks obtained from the AUTEC dataset~\cite{AUTEC}. These are compared to the standard deviation of the MPS for groups of five manually verified 4197 noise transients that we obtained on the Dominica Island. A description of the MPS calculation can be found in Section~\ref{sec:MPS} below. We find that the two groups of clicks and noise transients are well separated.}

\subsection{Identification of ROIs}

Our method for identifying ROIs comprises three steps. First, the signal is enhanced after a bandpass filter (BPF) for the desired frequency band of 2-24~kHz to increase the SNR using the TKEO operator~\cite{kaiser1990simple}, which has already been shown to be effective in processing SW clicks~\cite{lohrasbipeydeh2014adaptive},\cite{kandia2006detection},\cite{kandia2008phase}. The enhanced signal, $\mathbf{s}=[s_1,\ldots,s_N]^T$ is defined such that
\begin{equation}
	\begin{aligned}
		s_n= y_n^2-y_{n-1} \cdot y_{n+1}\quad, n=1,\ldots,N \;,
		\label{eq 2}
	\end{aligned}
\end{equation}
where $y_n$ is the sampled signal after the BPF. The process continues with a simple peak detector over $s_n$ to identify local energy maxima, and in terms of real-time processing, the number of identified transients can be limited (in our results, we considered up to 45 click).

\subsection{MPS evaluation}
\label{sec:MPS}

\begin{figure*}[]
 	\centerline{
 		\includegraphics[width=160mm,angle=0]{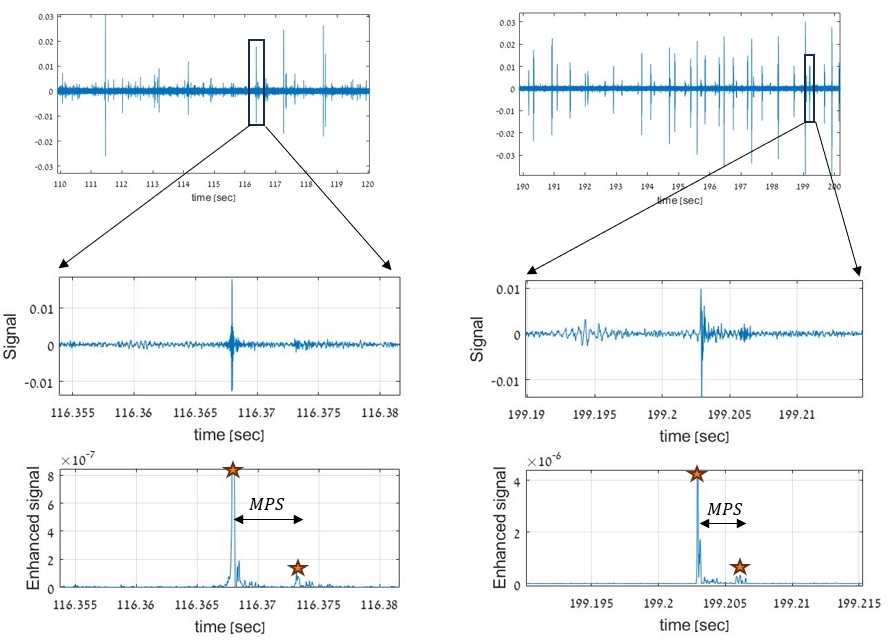}}
 	\caption{ Demonstration of the MPS calculation process for the case of a SW click (right panel) and for a transient noise (left panel). The top panels show recording buffers containing SW clicks and noise-only, respectively. The middle panels show the ROIs of a click and a noise transient. The lower blocks show the layout of the enhenced signals. The MPS is calculated based on the time difference between the two highest peaks of the enhanced signal (marked by orange stars).}
 	\label{MPS_examples}
 \end{figure*}

\remembertext{R1C15b}{We define the MPS as the time difference between the arrivals of the two most dominant (intense) pulses within the ROI. Let $t_{P_1}$ and $t_{P_2}$ be the arrival times of the highest and second highest peaks in the enhanced signal, $s_n$, respectively. The MPS is measured by
\begin{equation}
		\rho_{\mathrm{MPS}}= |t_{P_1}-t_{P_2}| \;.
		\label{eq MPS}
\end{equation}
Fig.~\ref{MPS_examples} shows the process of MPS calculation for the case of a SW click (right panel) and for a transient noise (left panel). In both cases, a similar multi-pulsed structure is observed. This result illustrates the challenge of classification based on single ROI.}
\begin{figure}[]
 	\centerline{
 		\includegraphics[width=90mm,angle=0]{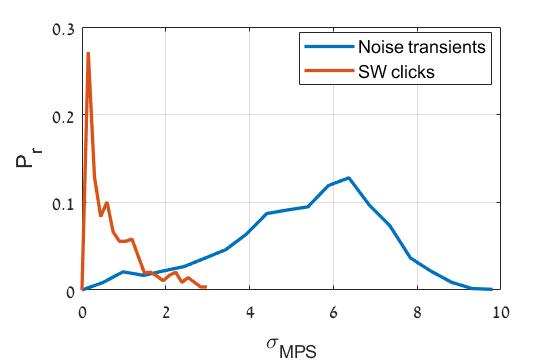}}
 	\caption{ MPS stability measurements evaluated in terms of the standard deviation of groups of 5 successive MPS measurements. }
 	\label{validation of MPS stability}
 \end{figure}
 
\begin{figure*}
\centering  
\begin{subfigure}[t]{.9\columnwidth}
\centering\includegraphics[width=.9\linewidth]{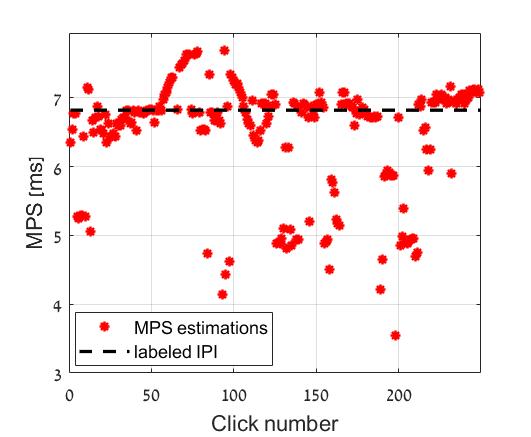}
    \caption{MPS estimation results over recording from the AUTEC database.}
  \end{subfigure}\quad
\begin{subfigure}[t]{.9\columnwidth}
\centering\includegraphics[width=0.95\columnwidth]{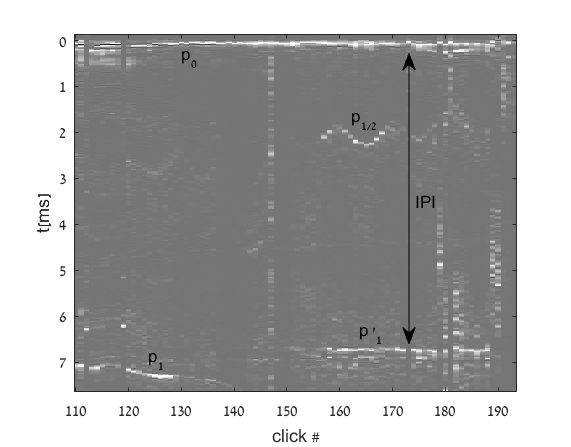}
    \caption{The normalized apparent level of 85 usual clicks emitted by A SW in the AUTEC recordings. The distance between the Type 1 pulses (i.e. $p_0$ and $p'_1$), indicated by the stable lines, determines the ground truth for the IPI assessment.}
  \end{subfigure}\quad
\caption{Demonstration of the relation between the MPS measure and the click's IPI.}
\label{IPI qulitative}
\end{figure*}

\remembertext{R1C12a}{Note that the MPS in \eqref{eq MPS} can capture not only the multipulse structure within the click, but also other reflections. In the first case, the MPS is expected to intersect with the IPI when the head of the whale is directed towards the receiver, while in other cases multipath reflections can be detected. Fig~\ref{IPI qulitative} shows the MPS measurements of the click series of a whale and the corresponding annotated pulses. It is obvious that the MPS measurements captured the IPI values for less than 50\% of the clicks. We thus cannot set a direct threshold for the MPS, by e.g. the maximum IPI value of a whale, but perform the detection based on the stability of the MPS.} \remembertext{R1C13}{Recall that we assume that the arrival times of the dominant multipath components remain stable along the investigated time buffer. Therefore, strong multipath affects all clicks in the examined buffer in a similar way, and the stability of the estimation and thus the performance of the detector is not significantly affected.}

\subsection{MPS Stability}\label{sec:clustering}

We assume that the time differences between the pulses of the clicks and the multipath components of a SW channel are approximately constant along the investigated time buffer. Therefore, we test the stability of the MPS as a detection measure. This stability can be measured by the MPS variance, which should approach zero to detect valid SW clicks. In practice, the variance can be affected by noise transients in the analyzed buffer or by emissions received in the same time buffer from two whales that differ in size, orientation or channel impulse response. To test the stability of a time series of MPS measured for a train of $M$ clicks, we propose instead a clustering scheme followed by a decision logic. 

From the set of $M$ clicks and the corresponding vector of MPS values, $\bar{\rho}^{\mathrm{MPS}}=\{\rho^{\mathrm{MPS}}_1,\ldots, \rho^{\mathrm{MPS}}_M\}$, our clustering maps MPS measures into M-dimensional binary assignment vectors, $\bf{c}_k=[\omega_{k,1},\ldots,\omega_{k,M}]\;, \quad , \ k=1\ldots K$, where $\omega_{k,l}=\{0,1\}$ is set to 1 if the $l$th MPS is assigned to the $k$th cluster. Let $\rho^{\mathrm{intense}}_{k,n}$ be the acoustic intensity of the click corresponding to the $n$th element in the cluster $\bf{c}_k$. In our clustering, we want to obtain a set of m-array clusters organized as a matrix $\bf{C}$ so that the MPS values in each cluster are fairly stable, as are the corresponding $\rho^{\mathrm{intense}}_{k,n}$ values. Note that not all MPS values need to belong to clusters to avoid outliers due to noise transients. Consequently, the number of clusters, $K$, is another clustering parameter that is limited by the maximum number of clustering possibilities,
\begin{equation}
    K_{\max}=\sum_{q=3}^{M}\frac{M!}{(M-q)!q!}\;.
    \label{eq: clustering possibilities}
\end{equation}
Further bounds are set for the ICI of each consecutive pair of clicks in the cluster to be within a minimum and maximum value, $\mathrm{ICI}_{\min}$ and $\mathrm{ICI}_{\max}$, respectively, and for the consistency of the ICI to be below a maximum value, $\mathcal{C}_\mathrm{max}$. As explained in \eqref{eq 1}, the consistency of the ICI for a triplet of consecutive MPSs, $t_i,t_{i+1},t_{i+2}$, is quantified as follows
\begin{equation}
 \mathcal{C}(t_i,t_{i+1},t_{i+2})=\left|\log \left( \frac{t_{i+2}-t_{i+1}}{t_{i+1}-t_i}\right)\right|\;,
\end{equation}
and we assign the arrival times of the click to a cluster, $\bf{c}_k$, by the vector $\bf{t}_k=[t_1,\ldots,t_{L_k}]$, where $L_k=\bf{c}_k \bf{c}_k^\mathrm{T}$.

We formalize the clustering solution to obtain $\hat{K}$ and $\hat{\bf{C}}$ by the following optimization problem to find unique clusters:
\begin{subequations}
\label{eq:clustering}
    \begin{align}
        &\hat{K}, \hat{\bf{C}}, = \argmin_{K, \bf{c}_k, \ k=1,\ldots,K}
             {\cal U}(K, \bf{c}_k)_{k=1,\ldots,K}&             \label{eq:clustering_a}\\
        &\textrm{s.t.} \quad \mathrm{ICI}_\mathrm{min} <t_{i+1}-t_i< \mathrm{ICI}_\mathrm{max}, \forall t_i,t_{i+1}\in\bf{t}_k \label{eq:clustering_b}&\\
                                  & \quad\quad \mathcal{C}(t_i,t_{i+1},t_{i+2})< \mathcal{C}_\mathrm{max}, \ \forall t_i,t_{i+1},t_{i+2}\in\bf{t}_k& \label{eq:clustering_c}\\ 
                                  & \quad\quad K<K_{\max}&\\
                                  &L_k\leq\rho_{\mathrm{click}}\label{eq:clustering_f}&\\
                                  &\quad\quad c_l \perp  c_k,~\forall l \neq k&
                                  \label{eq:clustering_e}\;,
    \end{align}
\end{subequations}
where 
\begin{eqnarray}
    {\cal U}(K, \bf{c}_k)_{k=1,\ldots,K} &=& \frac{1}{K}+\sum_{k=1}^K  \sigma(\bar{\rho}^{\mathrm{MPS}}\odot
             \bf{c}_k)+\alpha_1\exp^{-L_k}\nonumber\\
             &+&\alpha_2\left(1- \frac{( \sum_{n=1}^{L_k} \mathbf{\rho}_{\mathbf{k},n}^{\mathrm{intense}})^2}{L_k\cdot\sum_{n=1}^{L_k} \left(\mathbf{\rho}_{\mathbf{k},n}^{\mathrm{intense}}\right)^2}\right)
             \label{eq:utility}
\end{eqnarray}
$\sigma(\bf{x})$ refers to the standard deviation of the elements in the vector $\bf{x}$, $\odot$ refers to the Hadamard product and $\rho_{\mathrm{click}}$ is a threshold for the number of possible clicks in a cluster. The first term in the utility function \eqref{eq:utility} penalizes a small number of clusters; the second term favors clusters whose MPS elements are stable; the third term favors a higher rank for the clusters; and the fourth term penalizes clusters that contain clicks whose acoustic intensity is not balanced. Note that constraining the rank of the cluster in Constraint~\eqref{eq:clustering_f} allows distinguishing clusters that contain more than the expected number of transients. The latter is defined by an expectation for a minimum ICI value.
%$ICI(\mathbf{c}_k)=(t_m-t_l)\cdot\mathbf{c}_k,~ m>l,m=2,...,M,$ and $\mathcal{C}(\mathbf{c}_k)=\left|\log \left( \frac{t_q-t_m}{t_m-t_l}\right)\right|\cdot\mathbf{c}_k,~q>m>l,q=3,...,M,$ are vectors consisting the measured ICI and consistency of ICI within cluster $\mathbf{c}_k$, respectively, where $t_l,~ l=1,...,M$, are the transients' arrival time. 
In our results we follow \cite{miller2004sperm} and set
$\mathrm{ICI}_\mathrm{min}=0.4, \ \mathrm{ICI}_\mathrm{max}=2$ so that we set $\rho_{\mathrm{click}}=25$ for a 10~sec buffer. Following~\cite{baggenstoss2011separation}, we set $\mathcal{C}_\mathrm{max}=0.15$ for bound \eqref{eq:clustering_c}.

\subsection{Verification}\label{sec:verify}

\begin{figure}
	\centering
 \begin{subfigure}[t]{.95\columnwidth}    \centering\includegraphics[width=0.9\linewidth]{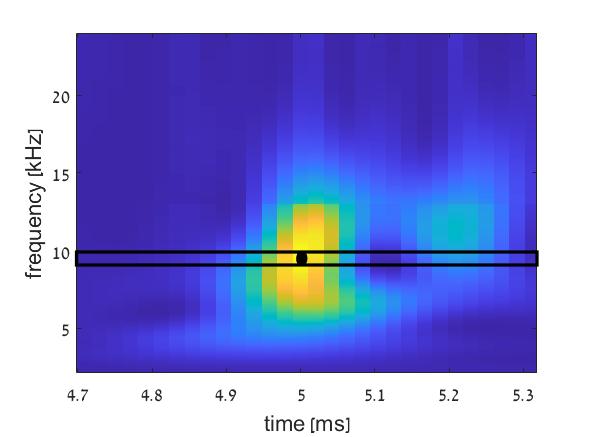}
  \caption{Super-resolution spectrogram of a pulse. The maximum frequency is marked in black dot.}
  \label{click spectrogram}
\end{subfigure}
  \begin{subfigure}[t]{.95\columnwidth}    \centering\includegraphics[width=0.9\linewidth]{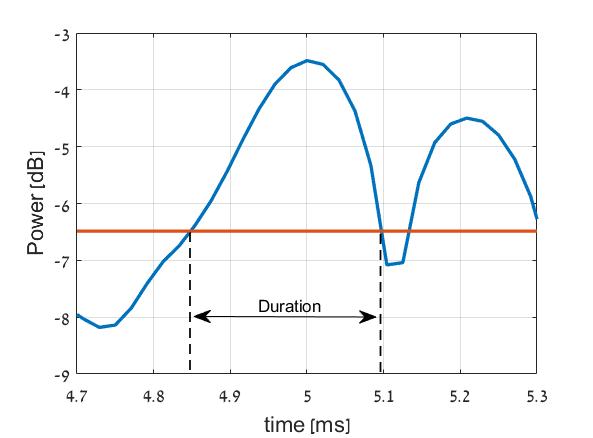}
    \caption{The power of the pulse across the dominant frequency. The pulse duration is computed by the 3dB cut-off time around the maximum power}
    \label{click duration}
\end{subfigure}
\caption{An example of the features extraction process for a single pulse.}
\label{Features extraction}
\end{figure}

\begin{figure}[]
 	\centerline{
 		\includegraphics[width=80mm,angle=0]{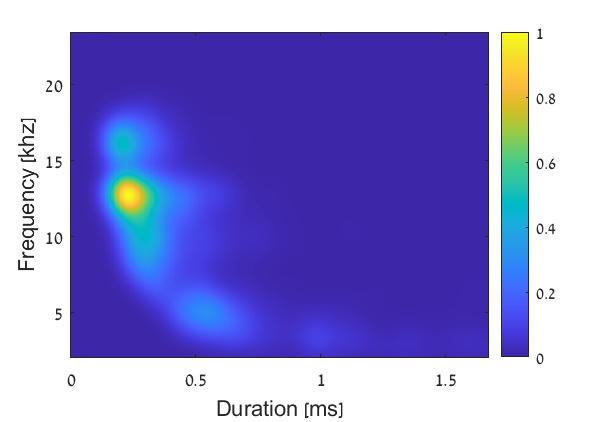}}
 	\caption{A layout of features' joint distribution obtained empirically over 5,200 annotated clicks from the
Dominica datase}
 	\label{Waveform Likelihood}
 \end{figure}

\remembertext{R1C14a}{Recall that the MPS values are obtained from the identified ROIs. These may contain noise transients that may happen to fulfill the constraints in \eqref{eq:clustering} and thus can be included in a valid clustering solution with potential impact on detection, or click emissions of a different whale species. For this reason, before determining the detection, we verify the clicks selected within each cluster against three conditions: 1) the MPS must not exceed a maximum value $\rho^{\mathrm{MPS,\max}}$, 2) the duration of the two dominant pulses selected for the MPS calculation must not exceed a threshold $d^{\max}$, and 3) the frequency response of the two dominant pulses should be feasible. Various features have been proposed in the literature for spectral analysis. Examples are the frequency profile~\cite{solsona2020detedit}, the spectral statistics~\cite{zaugg2010real,sanchez2010efficient} or the wavelet coefficient~\cite{van2009classification}. To avoid dependence on the recording device, whose frequency response may influence the measurements, we follow~\cite{lebien2018species} and use the most dominant frequency of the pulse, which is expected to be below a threshold $f^{\max}$, as the classification measure. The features are extracted for individual pulses and not for whole clicks to minimize tempo-spectral variability. For this purpose, we use a super-resolution spectrogram based on super-lets~\cite{moca2021time}. This tool offers a favorable compromise between spectral and temporal resolution compared to STFT spectrograms and allows the analysis of the inter-pulses of the clicks. %We calculate the signal's spectrogram over $10$ms time frames around the main peak arrival times (i.e., $t_{P_1}$, recall~\ref{sec:MPS}) of detected transients. Then, two features are extracted; the frequency peak and pulse duration. The duration is computed by the 3dB cut-off time around the maximum power. 
Fig.~\ref{Features extraction} shows an example of the feature extraction process for a single pulse.

%Our stability criteria handles the case of noise transients. Still, as demonstrated in Fig.~\ref{Stability results} some overlap exists. \textcolor{red}{Additionally, echolocation clicks from other marine species may yield stable MPS due to reflections from the sea-surface. With the aim of reducing the false alarm rate, in this section we present a verification stage based on individual click classification followed by a majority decision logic.  Verification is applied on the detected clicks $\hat{l}$. In case the contextual information is not available, e.g., in time buffers containing less than 6 clicks (often occurs for buffers at the beginning or at the end of a click train), each detected transient is verified directly.}

%Let $d_{k,n}$ and $f_{k,n}, / n=1\ldots L_k$ denote the duration and main frequency of the dominant pulse within the $n$th click associated with cluster $k$, respectively, and $\mathbf{\rho}_{\mathbf{k},l}^{\mathrm{dur}}$ and $\mathbf{\rho}_{\mathbf{k},l}^{\mathrm{freq}},\ l=1,\ldots,M_\mathbf{k}$ be accordingly the duration and frequency of the $l$'th click in cluster $\bf{c}_k$ . 
Threshold $\rho^{\mathrm{MPS,\max}}$ for the first condition is set according to the maximum delay spread assumed for the channel (40~msec in our results below). However, $d^{\max}$ and $f^{\max}$ are data-driven. Below are guidelines for setting these thresholds. Let $F_{\rho_d,\rho_f}$ be the joint probability density function of the pulse duration and the main frequency, respectively, for the legacy data of SW's validated clicks. For a target probability $P$, we set the values $d^{\max}$ and $f^{\max}$ by solving for
\begin{equation}
 P=\int_{d^{\max}}^\infty \int_{f^{\max}}^\infty F_{\rho_d,\rho_f}d\rho_dd\rho_f
\end{equation}
To obtain $F_{\rho_d,\rho_f}$, we numerically calculated the joint distribution for 5,200 annotated clicks from the Dominica dataset in this paper. The results in Fig.~\ref{Waveform Likelihood} show a distinct distribution}

% \textcolor{red}{Conventionally, the features are extracted from the entire apparent multi-pulsed content of the click. The duration of sperm whale clicks ranges from $0.1-20$ms  depending on axis angle of the measurement and the size of the animal~\cite{goold1995time,mohl2003monopulsed}. The energy of SW clicks is emphasized in the
% $2-15$kHz frequency ranges, depending also on the orientation of the hydrophone with respect to the animal~\cite{goold1995time,zimmer2005off}. To minimize the tempo-spectral variability, in this work we extract the features from individual pulses rather than from the entire click. To that end, we use a super-resolution spectrogram based on super-lets~\cite{moca2021time}. This tool provides a favorable trade-off between the spectral and temporal resolution compared to STFT spectrograms, allowing the analysis of click's inter-pulses. We calculate the signal's spectrogram over $10$ms time frames around the main peak arrival times (i.e., $t_{P_1}$, recall~\ref{sec:MPS}) of detected transients. Then, two features are extracted; the frequency peak and pulse duration. The duration is computed by the 3dB cut-off time around the maximum power. Fig.~\ref{Features extraction} shows a toy example of the features extraction over a SW click.}  

\iffalse
\textcolor{red}{\subsubsection{Decision logic} To reduce dependency on the training dataset, we verify the SW clusters using majority voting. Namely, a candidate cluster is verified if most of its transient components are classified as SW clicks. Formally, a decision regarding the validity of the candidate cluster $\mathbf{c}_k$ is determined according to the binary hypothesis test
\begin{equation}
\mathbf{c}_k\sim \begin{cases}
H_1,& \text{if} ~~\frac{1}{\delta}\displaystyle\sum_{k} g_{(\mathbf{X}_{\mathbf{c}_k})}\\
    H_{0},              & \text{otherwise}\;
\end{cases}
\end{equation}  where $H_0$ and $H_{1}$ are the null and alternative hypotheses representing a cluster of noise transients and a verified cluster of SW clicks, respectively, and
\begin{equation}
g_{(\mathbf{X}_{\mathbf{c}_k})} = \begin{cases}
1,& \text{if} ~~ \mathcal{L}_{(\mathbf{c}_k)}>\mathcal{L}_\mathrm{T}\\
    0,              & \text{otherwise}\;
\end{cases}
\end{equation}
  is a step function and $\mathcal{L}_{(T)}$ is a classification threshold.}
\fi

\remembertext{R1C14c}{For the clustering solution, ${\mathbf{\hat{c}}_k)}$, from \eqref{eq:clustering}, the result of the verification step are modified clusters, ${\mathbf{\tilde{c}}_k)}$, containing only MPS of validated clicks. The final recognition decision is based on their variance, as described below.}

\subsection{Detection Decision}\label{sec:detect_decision}

\remembertext{R1C14d}{Once the clustering is completed, the final recognition decision is based on the stability of the MPS clusters. Since we focus on presence detection in this paper, we use the utility function in \eqref{eq:utility} as a measure of stability, so that
\begin{equation}
\begin{cases}
H_1 \ (\mathrm{signal}),& \text{if}  \quad
{\cal U}(\hat{K}, \bf{\tilde{c}}_k)_{k=1,\ldots,\hat{K}}
< U_{\mathrm{T}}\;,\\
    H_{0} \ (\mathrm{noise}),              & \text{otherwise}\;,
\end{cases}
\label{eq:detection}
\end{equation}
where $U_{\mathrm{T}}$ is the only detection threshold used by our scheme. 
%where $H_0$ is the null hypothesis representing clusters associated with noise, and $H_{1}$ represents clusters comprised of SW's clicks, and $\sigma_{\mathrm{T}}$ is a detection threshold. The indices of the detected clicks are determined by 
%\begin{equation}
%   \hat{l}=\cup k,~~~ \forall \mathbf{c}_k \in  H_{1} 
%\end{equation}
}

\subsection{Complexity Analysis}\label{sec:complexity}

For the number of detected transients, $M$, and the maximum allowed number of transients in the cluster, $\rho_{\mathrm{click}}$, the complexity of our method is
\begin{equation} 
\mathcal{O}(M,\rho_{\mathrm{click}}) \propto \sum_{q=3}^{\rho_{\mathrm{click}}}\frac{M!}{(M-q)!q!}\;,%+\sum_{q=1}^{Q_O}\frac{Q!}{(Q-q)!q!}\;. 
\end{equation}
which reflects the number of clustering possibilities. %and the second term which stems from the orthogonality constraint in~\eqref{eq:clustering_e}, describes the number of possible cluster combinations dictated by the maximum number of orthogonal clusters, $Q_O$, and the number of candidate clusters, $Q$, that meet~\eqref{eq:clustering_b} and~\eqref{eq:clustering_c}. To allow fast processing, in this work we set $M=30$, $Q_R=7$ and adapt $Q_o$ such that
% \begin{equation} \max_{Q_O}
%              \sum_{q=1}^{Q_O}\frac{Q!}{(Q-q)!q!}<1\times10^6\;. 
% \end{equation}
At a setting of $M=30$, $\rho_{\mathrm{click}}=7$, processing takes less than 2~sec on an Intel i7 2.80GHz machine with an 11th generation processor with 16GB RAM.
%These settings lead to processing time of $<2$~sec per buffer in a standard laptop computer (Intel i7 2.80GHz 11th generation processor with 16GB of RAM).

\section{Experimental Results}
\label{results}
In this section, we examine the performance of our recognition system. We use an accumulated dataset consisting of roughly 40,000 SW clicks and over 302 hours of noise recordings obtained in three different marine environments. The SW data are a combination of recordings we collected in September 2023 approximately 2~km west off the coast of Dominica Island in the eastern Caribbean and 1,203 clicks observed by the Atlantic Undersea Test and Evaluation Center (AUTEC), which is available in~\cite{AUTEC}. \remembertext{R1C4c}{To collect the Dominica dataset, we used a self-built recorder equipped with an M36 Geospectrum hydrophone with near-constant sensitivity in the frequency range of 2~kHz to 20~kHz, sampling at 192~kHz. Recodings lasted for 5 days, in each the recorder was deployed at a depth of 20~m and was moored to a surface buoy that freely drifted. The water depth was about 1,600~m} The data from AUTEC was collected on the east coast of Andros Island, Bahamas, using a bottom-mounted hydrophone at a depth of approximately 2,000~m. Data collection was conducted by the Naval Undersea Warfare Center (NUWC), and more information about this deployment can be found at \cite{AUTEC}. Both datasets from Dominica and AUTEC were manually tagged by an expert from our team who is trained to identify SW clicks. To facilitate the tagging, the expert looked at the TKEO result to highlight acoustic pulse events, as well as the spectrogram display, and listened to the recordings.

\remembertext{R1C4e}{In addition to the click data, we also obtained a large data set consisting only of ambient noise, which was used for statistical analysis of false alarm rates. The data were collected from May to October 2018 at the shallow Themo observatory, located roughly 11~km west of the northern coast of Israel (see \cite{diamant2018themo}). Data were collected using an AMAR G3 acoustic recorder with two M36 Geospectrum hydrophones attached to the mooring at a depth of 90~m. Data were recorded at a sampling rate of 250~kHz per channel (3~bytes per sample). No SW has previously observed in the shallow THEMO observatory. This was verified by a sonar expert who manually tagged the entire data set and found that no SW echolocation clicks or codas had been recorded. This dataset reflects a very noisy marine environment with strong noise transients.}

\subsection{Benchmarks}\label{sec:benchmarks}

\remembertext{benchmark}{We compare our scheme with two benchmarks. The first one is called \textit{Statistical}~\cite{zaugg2010real} and is a classification approach based on high-order temporal and spectral features. The second benchmark is \textit{PamGaurd}, a multi-hypothesis tracking detector described in~\cite{macaulay2020passive}) and integrated into the open-source software package PamGuard~\cite{gillespie2009pamguard}. This approach comprises three stages: Pulse detection, ICI-based click-train detection and classification based on a spectral template. A detection threshold is set by the $\chi^2$ measure (calculated by Eq. 2.2, in~\cite{macaulay2020passive}), and the classification threshold is determined by a spectral correlation coefficient and set to 0.5. %The third benchmark is referred to as \textit{Triton}. This is a presence detector described in~\cite{frasier2021machine} that combines unsupervised clustering with machine learning based classification. The Triton detector uses 1) pulse detection, 2) unsupervised clustering and 3) network training and bin classification. The implementation of this method is available via the open source Triton software package~\cite{wiggins2010triton} using its SPICE detector, the Cluster tool and the NNet tool. 
We have chosen these detectors as benchmarks as they have only recently been introduced and are widely used.

The PamGaurd is an open-access code, while the 'Statistical' scheme was implemented by us according to the description in \cite{zaugg2010real}. For the latter, we extracted the nine statistical features from the temporal and spectral content of the click (equations (1)-(5) in \cite{zaugg2010real}). A feed-forward neural network was used as classifier, whose input consists of nine units corresponding to the nine features. A single hidden layer consisting of 25 sigmoid feature units was used, and the output layer consisted of a softmax neuron modeling the two target classes: SW clicks and noise. The network was trained with scaled conjugate gradient backpropagation. We used our manually annotated dataset of 29,902 samples from the Dominica dataset (14,829 SW clicks and 15,073 noise transients), split evenly at random into training ($70\%$), validation ($15\%$) and testing ($15\%$). A detection threshold is set via the output of the classifier (Eq.7 in~\cite{zaugg2010real}).}

\subsection{Performance evaluation}

\begin{figure}
	\centering
 \begin{subfigure}[t]{.95\columnwidth}    \centering\includegraphics[width=1\linewidth]{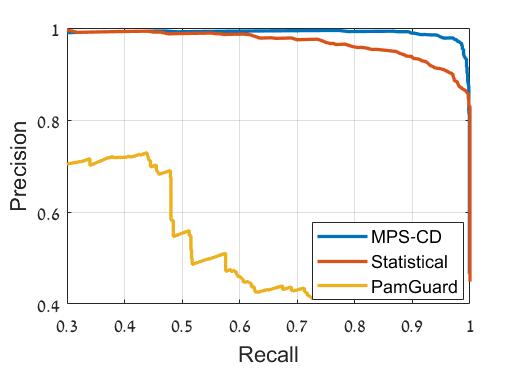}
  \caption{}
  \label{ROC Dominica}
\end{subfigure}
  \begin{subfigure}[t]{.95\columnwidth}    \centering\includegraphics[width=1\linewidth]{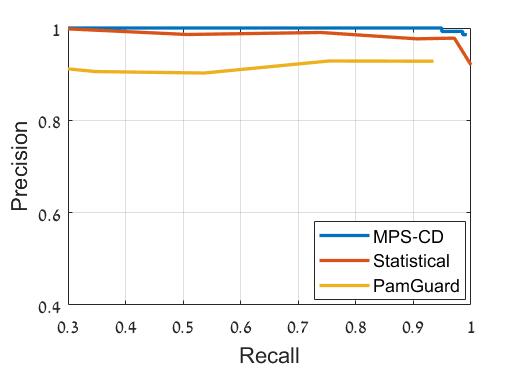}
    \caption{}
    \label{ROC AUTEC}
\end{subfigure}
\caption{Precision an recall curves evaluated over the Dominica (a) and the AUTEC (b) datasets.}
\label{ROC}
\end{figure}

\remembertext{R1C8b}{We investigate the performance of our scheme in terms of the detection performance for different marine environments with different noise conditions. While we rely on the MPS's stability, we avoid testing the accuracy of the MPS measure directly. This is because although the MPS can capture the IPI of the click (e.g., see Fig.~\ref{IPI qulitative}), it can also measure the difference between the two dominant multipath arrivals, and is therefore difficult to compare to a ground truth.} \remembertext{R1C4b}{We start by investigating the trade-off between the precision and recall defined by
\begin{equation}
 \mathrm{precision}=\frac{\mathrm{T}_\mathrm{p}}{\mathrm{T}_\mathrm{p}+\mathrm{F}_\mathrm{p}}, ~~ \mathrm{recall}=\frac{\mathrm{T}_\mathrm{p}}{\mathrm{T}_\mathrm{p}+\mathrm{F}_\mathrm{n}}\;,
\end{equation}
where $\mathrm{T}_\mathrm{p}$, $\mathrm{F}_\mathrm{p}$ and $\mathrm{F}_\mathrm{n}$ denote the rates for true positive, false positive and false negative, respectively. For all methods examined, the initial impulse threshold is set to detect at $\mathrm{SNR}=23dB$. The results are shown in Fig.~\ref{ROC Dominica} for the Dominica dataset and in Fig.~\ref{ROC AUTEC} for the AUTEC dataset. The noise-only buffers are identified in both cases manually both by spectrogram plots and by listening to the recordings. We identified 3.6 hours of pure noise-only for the Dominica dataset and 120~sec of noise for the AUTEC dataset. We observed dolphin clicks, ambient and ship noise, and other unidentified transients in the Dominica recordings, while the noise components in the AUTEC recordings consisted of ambient noise and creaks (see Sec.~\ref{sec: preliminaries}). We observe reasonable performance of PamGaurd for the AUTEC dataset but low performance for the Dominica dataset. This highlights PamGaurd sensitivity to noise transients. In contract, the statistical approach shows robustness for the marine environment in terms of the precision-recall trade-off, but, especially in the desired higher recall region, its performance is lower than that of MPS-CD.} 

% \begin{table}[]
% \caption{Percentage of false alarms. Results by thresholds set according to the AUTEC / Dominica datasets for a recall of 0.7, 0.75 and 0.9.}  % title of Table 
%  \centering      % used for centering table
% \begin{tabular}{lllllll}
% \hline
%                  & \multicolumn{2}{c}{Recall=0.7} & \multicolumn{2}{c}{Recall=0.75} & \multicolumn{2}{c}{Recall=0.9} \\ \hline
% Detector/Dataset & Dominica        & AUTEC        & Dominica       & AUTEC          & Dominica        & AUTEC        \\ \hline
% MPS-CD           & 1.16 / 0.33       & 0 / 0          & 1.16 / 0.33      & 0 / 0            & 6.06 / 0.83       & 0 / 0          \\
% Statistical      & 11.5 / 1.5        & 8.3 / 0        & 12.4 / 1.73      & 8.3 / 0          & 18.15 / 5         & 25 / 0         \\
% PamGaurd         & NA / 83            & NA / 83.3       & 5.3 /  92.7       & 66.6 / 83.3      & 7.2 / NA           & 83.3 / NA       \\ \hline
% \end{tabular}
% \end{table}

\begin{table}[]
\caption{Percentage of false alarms. Results by thresholds set according to the AUTEC / Dominica datasets for recall rates of 0.75 and 0.9.}  % title of Table 
 \centering      % used for centering table
\begin{tabular}{lllll}
\hline
                  & \multicolumn{2}{c}{Recall=0.75} & \multicolumn{2}{c}{Recall=0.9} \\ \hline
Detector/Dataset   & Dominica       & AUTEC          & Dominica        & AUTEC        \\ \hline
MPS-CD              & 1.16 / 0.33      & 0 / 0            & 6.06 / 0.83       & 0 / 0          \\
Statistical      & 12.4 / 1.73      & 8.3 / 0          & 18.15 / 5         & 25 / 0         \\
PamGaurd        & 5.3 /  92.7       & 66.6 / 83.3      & 7.2 / NA           & 83.3 / NA       \\ \hline
\end{tabular}
\label{Results- Pfa}
\end{table}

% \begin{table}[]
% \caption{False alarm rate per hour for the THEMO dataset. Results by thresholds set according to the ROC of the AUTEC / Dominica datasets in Fig.~\ref{ROC} for a recall of 0.7, 0.75 and 0.9.}  % title of Table 
% \centering 
% \begin{tabular}{lllll} \hline
% Detector/Operating point & Recall=0.7    & Recall=0.75   & Recall=0.9    &  \\ \hline
% MPS-CD                   & 0.35 / 0.066  & 0.37 / 0.075  & 5.25 / 0.25   &  \\
% Statistical              & 128.5 / 0.033 & 135.4 / 0.066 & 176.9 / 39.06 &  \\
% PamGaurd                 & NA / 33.12    & 3.96 / 35.64  & 10.44 / NA    & \\ \hline
% \end{tabular}
% \end{table}

\begin{table}[]
\caption{False alarm rate per hour for the THEMO dataset. Results by thresholds set according to the ROC of the AUTEC / Dominica datasets in Fig.~\ref{ROC} for a recall of 0.75 and 0.9.}  % title of Table 
\centering 
\begin{tabular}{llll} \hline
Detector/Operating point  & Recall=0.75   & Recall=0.9    &  \\ \hline
MPS-CD                   & 0.37 / 0.075  & 5.25 / 0.25   &  \\
Statistical             & 135.4 / 0.066 & 176.9 / 39.06 &  \\
PamGaurd                  & 3.96 / 35.64  & 10.44 / NA    & \\ \hline
\end{tabular}
\label{Results- FAR}
\end{table}

\remembertext{R1C4g}{Next, we investigate the sensitivity to different marine environments by analyzing the number of false alarms when analyzing the Dominica and AUTEC datasets separately. The results in Table~\ref{Results- Pfa} are given for two cases: when the threshold for detection is set by a target recall for the Dominica dataset (see Fig.~\ref{ROC Dominica}), and when the threshold is set by the same target recall for the AUTEC dataset (see Fig.~\ref{ROC AUTEC}). The recall values considered are 0.75 and 0.9\footnote{Since the PamGuard did not achieve a recall value of more than 0.75 in the Dominica dataset, we mark 'NA' for the target recall of 0.9.}. The large number of false alarms detected in PamGaurd for the AUTEC-based threshold is due to the existence of creak vocalizations, which are discarded by MPS-CD and Statistical, but not by PamGaurd. For the Dominica dataset, we find that the number of false alarms from MPS-CD is not significantly affected by the dataset for which the threshold was set, while a large change is observed for the two benchmarks. When we compare the percentage of false alarms between the datasets, we find a significant advantage of MPS-CD.}

\remembertext{R1C4a}{The performance for the THEMO dataset in setting the thresholds by the AUTEC/Dominica datasets is shown in Table~\ref{Results- FAR}. This dataset contains large amounts of noise-only buffers, allowing the false alarm rate per hour to be examined. The results show that the Statistical approach fails completely at a recall rate of 0.9 and shows large differences in the results between the AUTEC-based and Dominica-based thresholds at a recall rate of 0.75. PamGaurd shows better false alarm rates, but again there are large differences. In contrast, MPS-CD achieves low false alarm rates in all cases without the need for pre-training in this environment and with little variation between data sources for threshold setting.}

\subsection{Discussion}

\remembertext{R1C15g}{Our MPS-CD scheme is an unsupervised solution for the presence detection of SW clicks based on the stability of the MPS for clicks emitted by a SW. Through our analysis in Figs.~\ref{validation of MPS stability} and \ref{ROC} and Tables~\ref{Results- Pfa} and \ref{Results- FAR}, we have shown that using MPS stability as a detection metric has the advantage of operating well even in noisy marine environments and in the presence of multiple emitting SWs. In contrast to distribution or similarity analysis, stability testing does not require a large sample size, so detection can be performed for small time buffers of about 10~sec, as required for real-time applications.} %To further reduce complexity, the number of ROIs containing potential clicks can be limited identified clicks while limiting the number of concurrent emitting SWs by
%\begin{equation} \mathrm{Number~of~whales}=\frac{\mathrm{Number~of~allowed~transients}}{ \mathrm{ICI}\times\mathrm{Buffer~ length}}\;. 
%\end{equation}
%for example, for 45 ROIs and a 12~sec time buffer, the number of emitting SW our method can handle is between 2 to 9.}  

\remembertext{R1C14b}{We note that the verification process in Section~\ref{sec:verify} may be performed before the clustering in Section~\ref{sec:clustering} to remove invalid clicks before the MPS calculation. However, the effects of a possible error when removing a valid click are not the same if the verification is performed before clustering. The reason for this is the ICI restriction in \eqref{eq:clustering_b}. Specifically, if a valid click is accidentally eliminated from a sequence of clicks, the series of ICI measurements will show a large gap in the sequence of arrival times of the clicks, and the clustering solution will split the sequence into two groups, which may lead to disqualification of the entire series as the number of clicks is reduced. However, if a valid click is eliminated after clustering, this only affects the number of samples available for the final variance test in \eqref{eq:detection}.}

\section{Conclusion}\label{sec:conclude}

\remembertext{conclude}{In this paper, we presented MPS-CD: a method for detecting and classifying SW echolocation clicks in a realistic environment that includes noise transients and multiple emitting whales. Our method is based on the stability quantification of the MPS of the clicks, a measure of the time delay between the two most dominant pulses of the click. The MPS is calculated for each suspected click in a time buffer. To handle highly noisy environments, we cluster the MPS series into groups with consistent MPS values, satisfying constraints on ICI value and consistency. Verification aims to eliminate potential clicks that do not meet the pulse duration and center frequency limits and improves resilience to noise transients, and we provide guidelines for setting these limits. Performance is evaluated for three different marine environments, including seven months of noise-only data and approximately 15,000 manually labeled SW clicks; a dataset we share with the community along with our implementation code in~\cite{Dominica}. The scale of this data allows us to investigate the trade-off between precision and recall, as well as robustness to noisy marine environments. The results show that our approach achieves better detection results compared to two benchmarks, especially in noisy marine environments.} \remembertext{R1C18}{Future work will build on the MPS measure as well as statistical relationships between clicks to distinguish between emissions from different SWs.}

% \section*{Acknowledgments}
% The authors would like to thank Alik Chebotar, Shlomi Dahan and Liav Nagar for conducting the sea experiments, as well as Gill Danino for tagging the clicks. 

\ifCLASSOPTIONcaptionsoff
  \newpage
\fi



%\bibliographystyle{IEEEtran}
\bibliographystyle{unsrt}

\bibliography{Main}

%\newpage
%\input{Rebuttal}
%\input{Resubmission}

\end{document}